%% file: main.tex
\documentclass[aps,prl,twocolumn,groupedaddress]{revtex4-2}

\input{preamble}
\newcommand{\simplesection}[1]{{\par\it #1.---}}

\begin{document}

\title{Local Determinacy of Quantum Master Equations and a Mechanical Interpretation of the Multi-Mode Jaynes--Cummings and Central Spin Models}

\author{Jaeha Lee}
\affiliation{Institute of Industrial Science, The University of Tokyo, 5-1-5 Kashiwanoha, Kashiwa-shi, Chiba 277-8574, Japan}

\author{Miku Ishizaki}
\affiliation{Department of Electrical, Electronic, and Communication Engineering, Faculty of Science and Engineering, Chuo University, 1-13-27 Kasuga, Bunkyo-ku, Tokyo 112-8551, Japan}

\begin{abstract}
We propose a novel formulation of master equations for open systems wherein the evolution of a state is determined solely by its local behaviour at any point of time. Specifically, our formulation allows for a local interpretation of the workings of beyond-Markovian dynamics as opposed to the more common conception that non-Markovian state evolution is affected by its cumulative past history. Quite interestingly, local determinacy is found prevalent in quantum dynamics. We illustrate the advantages of our coordinate-free formulation with exact analyses on two physically relevant models.
\end{abstract}

\maketitle

\simplesection{Introduction}\label{sec:introduction}
Realistic quantum systems are seldom closed in the strict sense of word, as they are most typically under constant influence of the external environment.  Whereas the evolution of a closed quantum system is widely understood to follow unitary dynamics, that of an open system displays a genuinely different behaviour.  One of its most straightforward generalisations would arguably be the Markovian dynamics, whose precise definition has actually yet to come to a universal agreement~\cite{Wolf2008,WolfPRL2008,Breuer2009,Rivas2010,Chruscinski2011,Rivas2014,Breuer2016,Vega2017,Chruscinski2018}.  In the strongest sense, quantum Markovianity is perhaps most commonly equated with quantum dynamical semigroups, \textit{i.e.}, strongly continuous one-parameter semigroups of completely-positive and trace-preserving maps.  A useful characterisation of their infinitesimal generators was provided by Gorini, Kossakowski, and Sudarshan~\cite{Gorini1976} for finite-dimensional systems, and almost simultaneously by Lindblad~\cite{Lindblad1976} for the uniformly continuous families;  the autonomous (\textit{i.e.}, time-independent) first-order differential equation
\begin{equation}\label{def:semigroup_master_equation}
\rho^{\prime}(t)
	= -i\mathcal{A}\rho(t)
\end{equation}
associated with the said generator is now known as the Gorini--Kossakowski--Sudarshan--Lindblad (GKSL) master equation.  While the evolution of realistic quantum systems in general cannot be expected to possess a semigroup property, there are known situations in which approximation by quantum dynamical semigroups becomes reasonably justifiable (see, \textit{e.g.}, Ref.~\cite{Breuer2002}).

Evolutions beyond the Markovian regime are collectively referred to as non-Markovian dynamics.  
Among the efforts in formulating the general reduced dynamics of open quantum systems, the Nakajima--Zwanzig (NZ) master equation \cite{Nakajima1958, Zwanzig1960} stands out in generality and interpretationability.  In essence, the (autonomous) NZ master equation describes the evolution of quantum states in terms of the first-order linear integro-differential equation of the form
\begin{equation}\label{def:NZ_master_equation}
\rho^{\prime}(t)
	= -i \mathcal{B} \rho(t) + i \int_{0}^{t}\Phi(t - \tau) \rho(\tau)\, d\tau - i \varphi(t),
\end{equation}
wherein the future evolution is understood to be governed by a deterministic factor, the cumulative effect of its past trajectory, and an inhomogeneous term often ascribed to non-local initial correlation;  this allows for one of the most common interpretation of non-Markovianity that the evolution is affected by the former history (or `memory') of the states themselves.

We explore a novel class of the dynamics of open systems that covers a wide class of beyond-Markovian evolutions but without direct reference to the influence of its non-local past or future:  we say that a dynamics is \emph{locally deterministic} if the evolution of a state is determined solely by its `local behaviour' at some point of time.  For discrete dynamics, our local determinacy and traditional determinacy (\textit{i.e.}, Markovianity) become synonymous as each point of time is topologically isolated from the rest.  It is in the continuous regime that subtle, yet interesting distinction arises.

As we are familiar from Taylor's theorem, local properties of an analytic function may be fully encoded into the sequence of its higher-order derivatives.  In view of this, we formulate our master equation with differential equations of the form
\begin{equation}\label{def:local_master_equation_non-autonomous_coordinate}
\rho^{(n)}(t) = f\prn*{t; \rho(t), \rho^{\prime}(t), \ldots, \rho^{(m)}(t) },
\end{equation}
where $f$ is a smooth function, $\rho^{(k)}$ denotes the $k$-th derivative of a trajectory $\rho: t \mapsto \rho(t)$ representing the evolution of a quantum state, and $n > m$.  Our master equation~\eqref{def:local_master_equation_non-autonomous_coordinate} dictates that a higher-order derivative of a trajectory be uniquely specified by its `local behaviour' of lower order through the assignment of $f$; this assignment, as its effect cascades down the lower-order derivatives, eventually determines the successive infinitesimal step of its evolution.  Formally, this encompasses the GKSL equation in a natural way, as differential equations of the form~\eqref{def:semigroup_master_equation} may be understood as special instances of~\eqref{def:local_master_equation_non-autonomous_coordinate} for $n=1$.

A notable feature of our formulation is its coordinate-free formalism, which we will illustrate in this paper with examples;  as such, our master equation~\eqref{def:local_master_equation_non-autonomous_coordinate} is in general non-linear by design.  This is not only to introduce flexibility in our framework in comparison to the more traditional methods based on linear equations, but rather reflects a fundamental aspect of the notion of `higher-order differentiations': they have in general no invariant linear structure, as we may infer from the fact that accelerations behave differently from velocities or differentials of functions in classical (analytical) mechanics.  A formal mathematical discourse is beyond the scope of this paper, and thus will be given elsewhere.

In this paper, we will illustrate the significance and advantages of our formulation with two physically relevant examples of non-Markovian dynamics, for which---all the more so by the unbounded nature of the total Hamiltonians---exact analyses based on the traditional master equations would be challenging, if not close to unmanageable.

\simplesection{Scope of Local Determinacy}
We first demonstrate that every reduced quantum dynamics is in fact locally deterministic as long as the Hamiltonian $H$ of the total system is bounded;  this guarantees the prevalence of the class.

In presenting our claim, we first prepare symbols for the Liouville operator $\mathcal{L}_{H} : X \mapsto [H,X]$ associated with $H$ and the one-parameter group $\mathcal{U}_{H}(t) \defeq \exp{\prn{-i\mathcal{L}_{H}t}}$ generated by it.  With these, we introduce the reduced trajectory
\begin{equation}\label{def:reduced_trajectory}
\rho(t)
	\defeq \mathrm{Tr}_{E}\brk*{\,\mathcal{U}_{H}(t)\rho_{\mathrm{tot}}}
\end{equation}
of open quantum systems, where $\rho_{\mathrm{tot}}$ is the initial state of the total system and $\mathrm{Tr}_{E}\brk*{\,\cdot\,}$ denotes the partial trace over the environment.  We have the following theorem: if any two reduced trajectories match $\rho_{1}(t_{k}) = \rho_{2}(t_{k})$ on some convergent sequence $(t_{k})_{k\in\mathbb{N}}$, then $\rho_{1}(t) = \rho_{2}(t)$ is globally identical.  More specifically and colloquially, if any two trajectories behave the same way in an infinitesimal neighbourhood of a point of time $t$, then they behave the same for all time.

A sketch of proof would be as follows:  the uniform continuity of $\mathcal{U}_{H}(t)$ entails its analyticity, which, combined with the continuity of the (partial) trace, points to the analyticity of the reduced trajectory $\rho(t) = \sum_{k=0}^{\infty} t^{k} \mathrm{Tr}_{E}\brk{\,(-i\mathcal{L}_{H})^{k}\rho_{\mathrm{tot}}}/k!$ with the series being strongly convergent.  The rest follows from the identity theorem.  We note that our result may be straightforwardly generalized to any reduced dynamics of uniformly continuous $C_{0}$-semigroups.

\simplesection{Existence of Master Equations}
One would be naturally interested in the existence of differential equations (of finite order) of the form~\eqref{def:local_master_equation_non-autonomous_coordinate} in accordance with the above theorem; in fact, we have a positive result whenever the total Hamiltonian has finite point spectrum (\textit{i.e.}, eigenvalues) and the system of interest is of finite dimension.

To see this, we may for instance examine the reduced trajectories~\eqref{def:reduced_trajectory} in the Laplace domain to seek for a `rational' expression
\begin{equation}\label{def:reduced_trajectory_Laplace}
\prn{\mathcal{L}\rho}(s)
	\defeq \int_{0}^{\infty} \rho(t)e^{-st}\, dt
	= P(s)^{-1} Q(s; \rho_{\mathrm{tot}}) 
\end{equation}
with a monic polynomial operator $P(x) = x^{n} + \sum_{k=0}^{n-1} P_{k}x^{k}$ and another polynomial operator $Q(x; \rho_{\mathrm{tot}})$ of lower order that depends on $\rho_{\mathrm{tot}}$.  We note that such an expression is always available as we may choose $P(x) = \prod_{k=1}^{n} \prn{x + i \lambda_{k}}$ to be the minimal polynomial of the generator $\mathcal{L}_{H}$ with $\brc{ \lambda_{1}, \ldots, \lambda_{n}}$ denoting its eigenvalues;  this may be observed by recalling that the Laplace transform of a contraction semigroup is the resolvent mapping of its generator, or more specifically $\prn{\mathcal{L}\,\mathcal{U}_{H}}(s) = \prn{s - \mathcal{L}_{H}}^{-1}$ for our case.

Under these assumptions, we claim the following: the reduced trajectories~\eqref{def:reduced_trajectory} are solutions to the differential equation
\begin{equation}\label{def:reduced_trajectory_differential_equation}
\rho^{(n)}(t) = - \sum_{k=0}^{n-1}P_{k}\rho^{(k)}(t)
\end{equation}
associated with $P$.  Note the non-uniqueness of such a differential equation~\eqref{def:reduced_trajectory_differential_equation} in accordance with the non-uniqueness of the expression~\eqref{def:reduced_trajectory_Laplace};  for a compact description, one would be interested in aiming for an `irreducible' expression.   

The general solutions to the linear differential equation~\eqref{def:reduced_trajectory_differential_equation} are found to be uniquely specified by the tuples of their initial values $\rho_{n-1}(0) \defeq \prn{ \rho(0),\rho^{\prime}(0),\ldots,\rho^{(n-1)}(0) }$.  Our result specifically entails that each state $\rho_{\mathrm{tot}}$ of the total system---be it initially correlated or not---translates to a tuple of initial values $\rho_{n-1}(0)$ in such a way that the reduced trajectory generated by the former may be represented as the unique solution associated with the latter;  formally, we introduce the parametrized mapping
\begin{equation}\label{def:flow_polynomial}
\rho_{n-1}(0) \mapsto \vartheta_{P}\prn{t; \rho_{n-1}(0)} = \rho(t)
\end{equation}
that maps an initial value to its unique solution at time $t$ and say that the assignment~\eqref{def:reduced_trajectory} of the initial states to the reduced trajectories
\begin{equation}\label{eq:polynomial_equation_flow_factorisation}
\mathrm{Tr}_{E}\brk*{\,\mathcal{U}_{H}(t)\rho_{\mathrm{tot}}} = \vartheta_{P}\prn*{t; \pi_{P}\prn*{\rho_{\mathrm{tot}}}}
\end{equation}
factors through $\vartheta_{P}$ to yield a unique linear map $\pi_{P} : \rho_{\mathrm{tot}} \mapsto \rho_{n-1}(0)$ with the said characterisation.

As for the sketch of proof, we first find that the map~\eqref{def:flow_polynomial} in the Laplace domain admits a `rational' expression
\begin{equation}\label{eq:polynomial_equation_flow_Laplace}
P(s)\prn*{\mathcal{L}\vartheta_{P}}\prn*{s;\rho_{n-1}(0)} = \sum_{k=1}^{n}\sum_{i=1}^{k}s^{k-i}P_{k}\rho^{i-1}(0)
\end{equation}
with the convention $P_{n}=\id$, which may be formally observed by applying the formula $\prn*{\mathcal{L}\rho^{(n)}}(s) = s^{n}\prn{\mathcal{L}\rho}(s) - \sum_{k=1}^{n}s^{n-k}\rho^{k-1}(0)$ to the differential equation~\eqref{def:reduced_trajectory_differential_equation}.  We here find the injection $\rho_{n-1}(0) \mapsto P(s)\prn{\mathcal{L}\vartheta_{P}}\prn{s;\rho_{n-1}(0)}$ to be a bijection if understood as a linear map into the space of polynomial operators of order $n-1$, as the domain and the codomain are of the same finite dimension.  We may combine this observation with \eqref{def:reduced_trajectory_Laplace} and \eqref{eq:polynomial_equation_flow_Laplace} to conclude the unique existence of the aforementioned map $\pi_{P}$ satisfying~\eqref{eq:polynomial_equation_flow_factorisation}.

\simplesection{Jaynes--Cummings Model on Resonance}\label{sec:JC_on_resonance}
For our first demonstration, let us turn to the multi-mode JC model, which concerns a two-level quantum system centred around an infinite number of quantum harmonic oscillators prepared in the vacuum states;  the evolution of the total system is governed by the Hamiltonian
\begin{equation}\label{eq:JC_Hamiltonian}
H =
	\frac{\omega_{S}}{2}\sigma_{+} \sigma_{-} + \sum_{k=-\infty}^{\infty} \prn*{ g_{k}\sigma_{+}a_{k} + g_{k}^{\ast}\sigma_{-}a_{k}^{\dagger} + \frac{\omega_k}{2}a_{k}^{\dagger} a_{k} }
\end{equation}
with the ladder operators $\sigma_{\pm} \defeq \sigma_{1} \pm i \sigma_{2}$ on the central system, the creation $a_{k}^{\dagger}$ and annihilation $a_{k}$ operators on the $k$-th quantum harmonic oscillator, the coupling constants $g_{k}$, and the angular frequencies $\omega_{S}$, $\omega_{k}$.

For our purpose, we examine the spectral density of the Lorentzian form $J(\omega) = (2\pi)^{-1} \gamma_{0} \lambda^{2} /  \prn{ \prn{\omega_S - \omega}^{2} + \lambda^{2} }$, which is known (see, \textit{e.g.}, Ref.~\cite{Breuer2002}) to allow for a solvable solution; the Bloch-vector representation of the reduced trajectories~\eqref{def:reduced_trajectory} read
\begin{equation}\label{eq:JC_trajectory}
\vect{r}(t)
	= \diag{\brk*{g_{\epsilon}(t),g_{\epsilon}(t),g_{\epsilon}(t)^{2}}}\,\vect{r} + \prn*{0, 0, g_\epsilon(t)^{2}-1}
\end{equation}
for that of the initial state $\vect{r} = (r_{1},r_{2},r_{3}) \in \mathbb{R}^{3}$, $\norm{\vect{r}} \leq 1$, where $g_{\epsilon}(t) \defeq e^{-\lambda t/2} \prn{ \cosh{\prn{\epsilon t/2}} + (\lambda/\epsilon) \sinh{\prn{\epsilon t/2}} }$ with the shorthand $\epsilon^{2} \defeq \lambda (\lambda - 2 \gamma_0 )$;  for the critical parameter $\epsilon = 0$, we adopt $g_{0}(t) \defeq e^{-\lambda t /2} \prn*{ 1 + \lambda t/2 } = \lim_{\epsilon \to 0}{g_{\epsilon}(t)}$ as its definition.

In general, the family of curves~\eqref{eq:JC_trajectory} cannot be described as solutions to a first-order---be it autonomous or non-autonomous---differential equation, as multiple curves intersect at time $t$ satisfying $g_{\epsilon}(t)=0$;  this indeed becomes relevant when $\epsilon^{2} < 0 \iff 2\gamma_{0} > \lambda$.  In other words, an exact master equation of the (time-local) GKSL form in general fails to exist for this model.

In fact, we may apply our method introduced above to reveal a differential equation of the polynomial form~\eqref{def:reduced_trajectory_differential_equation} of order $3$ (but no less).  This implies the local determinacy of the family of curves~\eqref{eq:JC_trajectory} despite the non-Markovian nature of the dynamics.  
We hereby seek further and propose our master equation of order $2$;  this not only attains the most compact description, but also offers interesting insights in relation to classical mechanics.

To illustrate our point, we may for instance make use of the diffeomorphism
\begin{equation}
\varphi(x,y,z) = \prn*{ x, y, (1 + z)/\prn*{x^{2} + y^{2}} }
\end{equation}
on $\mathbb{R}^{2}_{\times}\times\mathbb{R}$ as the the smooth chart of our choice, where $\mathbb{R}^{2}_{\times} \defeq \mathbb{R}^{2}\setminus\{0\}$ denotes the punctured Euclidean plane; simple computation reveals the coordinate representation 
\begin{equation}\label{eq:JC_trajectory_coordinate}
\vect{\gamma}(t)
	\defeq \varphi(\vect{r}(t))
	=\prn*{ r_{1} g_{\epsilon}(t),\, r_{2} g_{\epsilon}(t),\, \frac{1 + r_{3}}{ r_{1}^{2} + r_{2}^{2}} }
\end{equation}
of the curves~\eqref{eq:JC_trajectory}.  We may apply our aforementioned method to reveal an autonomous second-order differential equation of the polynomial form~\eqref{def:reduced_trajectory_differential_equation} reading
\begin{equation}\label{eq:JC_master_equation_coordinate}
\ddot{\vect{\gamma}}(t) = - \frac{1}{4} \prn*{ \lambda^{2} - \epsilon^{2} } Q_{3}\vect{\gamma}(t) - \lambda Q_{3}\dot{\vect{\gamma}}(t)
\end{equation}
in our local coordinates; here, $Q_{i} \defeq I - P_{i}$ denotes the complement of the orthogonal projection $P_{i}\vect{e}_{j} \defeq \vect{e}_{j}\delta_{ij}$ on the $i$-th axis with $\vect{e}_{i}$ being the $i$-th element of the natural basis, $i=1,2,3$.
We here find the dynamics of the model to be of the same nature as that of the familiar Newtonian equation of motion for the classical $2$-dimensional isotropic damped harmonic oscillator.

We may compute the transition functions associated with the coordinate transformation to arrive at our master equation
\begin{equation}\label{eq:JC_master_equation}
\rho^{\prime\prime}(t) = f_{\epsilon}\prn{\rho(t), \rho^{\prime}(t)}
\end{equation}
in explicit density-matrix formulation with the smooth (non-linear) function
\begin{multline}\label{eq:JC_reference_function}
f_{\epsilon}\prn*{\xi,\upsilon}
	\defeq - \frac{1}{4}\prn*{ \lambda^{2} - \epsilon^{2} } \mathcal{Q}_{3}\xi - \lambda \mathcal{Q}_{3}\upsilon \\
		+ 2\biggl( - \frac{1}{4} \prn*{\lambda^{2}-\epsilon^{2}} - \frac{\lambda}{2} h_{3}\prn*{\xi,\upsilon} - h_{3}\prn*{\xi,v}^{2} \\ + g_{3}\prn{\xi,\upsilon}^{2} \biggr) \prn*{\frac{\sigma_{3}}{2} + \mathcal{P}_{3}\xi} + 2 h_{3}\prn*{\xi,\upsilon} \mathcal{P}_{3}\upsilon
\end{multline}
defined over operators $\xi$ and $\upsilon$. Here, we introduced the projections $\mathcal{P}_{i} \xi \defeq \sigma_{i} \tr{\brk*{\sigma_{i}\xi}}/2$, $\mathcal{Q}_{i} \xi\defeq \prn*{\xi - \sigma_{i}\xi\sigma_{i}}/2$ and smooth functions $g_{i}\prn{\xi,\upsilon} \defeq \norm{\mathcal{Q}_{i}\upsilon}/\norm{\mathcal{Q}_{i}\xi}$, $h_{i}\prn{\xi,\upsilon} \defeq 2 \inpr{\xi}{\mathcal{Q}_{i}\upsilon}/\norm{\mathcal{Q}_{i}\xi}^{2}$, $i = 1,2,3$, with $\norm{\xi}^{2} \defeq \tr{\brk*{\xi^{*}\xi}}$ and $\inpr{\xi}{\upsilon} \defeq \tr{\brk*{\xi^{*}\upsilon}}$ respectively denoting the Hilbert--Schmidt norm and the associated inner product.  Despite its apparent complexity, our master equation~\eqref{eq:JC_master_equation} retains its essence under coordinate transformations: it is `isomorphic' to the second-order differential equation~\eqref{eq:JC_master_equation_coordinate} of a well-known type we are familiar with in elementary classical mechanics. 

Let us close our first demonstration with a remark on the initial value problem: 
we are interested in characterizing the solution space of our (non-linear) master equation~\eqref{eq:JC_master_equation}.  For this, we may take advantage of the coordinate representation~\eqref{eq:JC_master_equation_coordinate} to pin down the map~\eqref{def:flow_polynomial} representing its general solutions, which may be subsequently transformed into the desired density-matrix expression.  Straightforward computation and transformation of~\eqref{def:flow_polynomial} reveals:  there exists a one-to-one correspondence between the general solutions
\begin{equation}\label{eq:JC_flow}
\rho(t) = \vartheta_{\epsilon}\prn*{t; \rho(0), \rho^{\prime}(0)}
\end{equation}
to our master equation~\eqref{eq:JC_master_equation} and their initial values $\rho_{1}(0) \defeq \prn*{\rho(0), \rho^{\prime}(0)}$, where
\begin{multline}
\vartheta_{\epsilon}\prn*{t; \xi, \upsilon} \defeq
	 \mathcal{P}_{0} \prn*{ \xi + \upsilon t }
		+ \mathcal{Q}_{3} \varphi_{\epsilon}\prn*{t;\xi, \upsilon} \\
		+ g_{3}\prn*{ \xi, \varphi_{\epsilon}\prn*{t;\xi, \upsilon} }
			\bigl( \prn*{ 1 - h_{3}(\xi,\upsilon) t } \\ \times \prn*{ \frac{\sigma_{3}}{2} + \mathcal{P}_{3} \xi } + \mathcal{P}_{3}\upsilon t \bigr) - \frac{\sigma_{3}}{2}
\end{multline}
with the function $\varphi_{\epsilon}\prn{t;x, v} \defeq e^{-\lambda t /2} \prn{ \cosh{\prn{\epsilon t/2}} \mathcal{Q}_{3}x + \prn{2/\epsilon} \sinh{\prn{\epsilon t/2}} \prn{ \prn{\lambda/2} \mathcal{Q}_{3}x + \mathcal{Q}_{3}v } + \mathcal{P}_{3}\prn{x + tv}}$.  Here, $\mathcal{P}_{0} \xi \defeq \sigma_{0} \tr{\brk*{\sigma_{0}\xi}}/2$ wheras all the other symbols are as in \eqref{eq:JC_reference_function}.  We may confirm that the trajectories~\eqref{eq:JC_flow} are indeed solutions to our master equation~\eqref{eq:JC_master_equation} through direct computation.  The reduced trajectories represented by~\eqref{eq:JC_trajectory} are precisely the solutions corresponding to the initial values without `initial velocity': $\rho^{\prime}(0) = 0$.

\simplesection{Infinite Central Spin System}\label{sec:central_spin}
For our second demonstration, let us consider a two-level quantum system centred around an infinite number of arbitrary external systems.  We consider the total Hamiltonian of the form
\begin{equation}\label{eq:central_spin_Hamiltonian}
H
	= \omega S + \lim_{n\to\infty} \sum_{k=-n}^{n} \frac{S \otimes B_{k}}{\sqrt{2n+1}} + \sum_{k=-\infty}^{\infty} H_{k}
\end{equation}
specified by a traceless and normalized reference observable $S = \vect{s} \cdot \vect{\sigma}$, $\norm{\vect{s}} = 1$, of the central system, the free Hamiltonian $H_{k}$ of each 
subsystem of the environment satisfying $[H,H_{k}] = 0 \iff [H,B_{k}] = 0$, and the angular frequency $\omega \in \mathbb{R}$.
The initial state $\rho_{\mathrm{tot}} = \rho_{S} \otimes \rho_{E}$ is prepared in such a way that the distributions of the external observables $B_{k}$ on the state $\rho_{E}$ are independent and identically distributed with vanishing mean $\mu=0$ and non-vanishing but finite standard deviation $0 < \sigma < \infty$.

Under this universal setup, we find through application of the classical central limit theorem the Bloch-vector representation
\begin{equation}\label{eq:CS_trajectory}
\vect{r}(t)
	= P_{\vect{s}}\vect{r} + e^{-2\sigma^{2}t^{2}} \prn*{ \cos{(2\omega t)} Q_{\vect{s}} 
		+ \sin{(2\omega t)} J_{\vect{s}} }\, \vect{r}
\end{equation}
of the reduced trajectories~\eqref{def:reduced_trajectory} of our model, where we introduce the orthogonal projection $P_{\vect{s}} \defeq \vect{s} \otimes \vect{s}^{t}$ on the linear span of $\vect{s}$, its complement $Q_{\vect{s}} = I - P_{\vect{s}}$, and $J_{\vect{s}} : \vect{r} \mapsto \vect{s} \times \vect{r}$.  At this point, we find that the curves~\eqref{eq:CS_trajectory} are never solutions to a differential equation of the polynomial form~\eqref{def:reduced_trajectory_differential_equation} of finite order.  In spite of this, we demonstrate below that the dynamics in fact obeys an autonomous second-order differential equation.

In constructing our coordinate chart, we introduce the pullback $\Log{(x,y)} \defeq \prn{\log{\sqrt{x^{2}+y^{2}}},\ \arg{\prn{x+iy}}}$ of (the principal value of) the complex logarithmic function on the punctured Euclidean plane $\mathbb{R}^{2}_{\times}$ in view of the usual identification $\mathbb{R}^{2} \simeq \mathbb{C}$ as real vector spaces.  For visual ease, we also make use of a rotation matrix $R \in \operatorname{SO}(3)$ satisfying $R\vect{e}_{1} = \vect{s}$ and another $S \in \operatorname{SO}(3)$ specified by $S\vect{e}_{1} = \vect{e}_{2}$, $S\vect{e}_{2} = \vect{e}_{3}$, and $S\vect{e}_{3} = \vect{e}_{1}$.  With these, we hereby introduce the coordinate map
\begin{equation}
\varphi = S \circ \prn{\operatorname{id}\times\Log} \circ R^{-1}
\end{equation}
of our choice, where $\operatorname{id}(x) \defeq x$ denotes the identity mapping on $\mathbb{R}$;   simple observation reveals the coordinate representation
\begin{equation}\label{eq:CS_trajectory_coordinate}
\vect{\gamma}(t)
	= \prn*{2 \omega t, 0, -2\sigma^{2}t^{2}} + \vect{\gamma}(0)
\end{equation}
of the curves~\eqref{eq:CS_trajectory}, where the identity is understood to hold modulo $(-\pi,\pi]$.  Here, the initial value is $\vect{\gamma}(0)
	= \prn*{ \arg{\prn*{ a_{1}(\vect{r}) + i a_{2}(\vect{r}) }},\, \vect{s}\cdot\vect{r},\, \log{ \norm*{Q_{\vect{s}}\vect{r}} } }$
with $a_{i}(\vect{r}) \defeq \prn*{R\vect{e}_{i+1}} \cdot \vect{r}$ defined for $i = 1,2$; note that the freedom of choice of $R$ only affects the initial values.  At this point, we immediately realize that the curves~\eqref{eq:CS_trajectory_coordinate} are solutions to the autonomous second-order differential equation reading
\begin{equation}\label{eq:CS_master_equation_coordinate}
\ddot{\vect{\gamma}}(t) = - 4\sigma^{2} \vect{e}_{3}
\end{equation}
in local coordinates;  this in essence is no different than the classical Newtonian equation of motion of a point mass under a uniform gravity field.

In explicit density-matrix formulation, the second-order differential equation~\eqref{eq:CS_master_equation_coordinate} translates to our master equation
\begin{multline}\label{eq:CS_master_equation}
\rho^{\prime\prime}(t)
	= - 4\sigma^{2} \mathcal{Q}_{S}\rho(t) \\ + g_{S}\prn*{\rho(t),\rho^{\prime}(t)}\, \mathcal{R}_{S}\brk*{\rho(t),\rho^{\prime}(t)}\, \rho^{\prime}(t),
\end{multline}
where we have introduced the shorthands for the projection $\mathcal{Q}_{S} \xi \defeq \prn{\xi - S \xi S}/2$, the function $g_{S}\prn{\xi,\upsilon} \defeq \norm{\mathcal{Q}_{S}\upsilon}/\norm{\mathcal{Q}_{S}\xi}$, and the map $\mathcal{R}_{S}\brk{\xi,\upsilon} \defeq \mathcal{Q}_{S} + \mathcal{K}_{S}\brk{\xi,\upsilon} + \prn{ 1 + m_{S}\prn{\xi,\upsilon} } \mathcal{K}_{S}\brk{\xi,\upsilon}^{2}$, which is in turn defined by the map $\mathcal{K}_{S}\brk{\xi,\upsilon} : \tau \mapsto - \comm{ \comm{\mathcal{Q}_{S}\xi}{\mathcal{Q}_{S}\upsilon} }{ \mathcal{Q}_{S} \tau } /2$ and  the function $m_{S}\prn{\xi,\upsilon} \defeq \inpr{\xi}{\mathcal{Q}_{S}\upsilon} / \norm{\mathcal{Q}_{S}\xi}\norm{\mathcal{Q}_{S}\upsilon}$.  The initial value problem may be also accounted for in a parallel manner as in the first model.

\simplesection{Mechanical Interpretation}
The reduced quantum dynamics of our examples are found to be structurally of the same nature as classical systems.  Our findings specifically allow for a mechanical interpretation of the systems.  For a quick look on some of its implications, let us take the second system as an example.  

Being a `conservative' system unlike the `dissipative' JC counterpart, the `equation of motion' of the central spin system admits a simple description; in fact, our master equation~\eqref{eq:CS_master_equation} may also be recovered from the standard Euler--Lagrange equation $\partial L/\partial \vect{q} - \prn{d/dt}\prn{\partial L/\partial \dot{\vect{q}}_{i}} = 0$ under the Lagrangian reading $L(\varphi,\dot{\varphi}) = \dot{\varphi}_{3}^{2}/2 - 4\sigma^{2}\varphi_{3}$ in our coordinates (with unit `mass' $m=1$).  We may also turn to Hamilton's equations $d\vect{q}/dt = \partial H/\partial \vect{p}$, $d\vect{p}/dt = - \partial H/\partial \vect{q}$ to obtain the same result with the classical Hamiltonian reading
\begin{equation}
H\prn*{\varphi,\psi} \defeq \frac{1}{2}\psi^{2} + 4\sigma^{2}\varphi^{2},
\end{equation}
where $\psi = \partial\mathcal{L}/\partial\dot{\varphi}$ is the conjugate momentum.  Structurally, the environment manifests itself as a `conservative force field' that is external to the freely moving particle representing the system of interest, and its effect is proportional to the property (\textit{i.e.}, variance of the distribution) of the environmental observables.  The angular frequency $\omega$ translates as the initial velocity of the particle, thereby contributing to the `energy' of the open system.

\begin{acknowledgments}
This work was supported by JSPS Grant-in-Aid for Scientific Research (KAKENHI), Grant No.~JP22K13970, JST Moonshot, Grant No.~JPMJMS226C, and CREST, Grant No.~JPMJCR2315.
\end{acknowledgments}

\bibliography{main}

\end{document}

%% file: preamble.tex
\usepackage{amsmath}
\usepackage{amsfonts}
\usepackage{amssymb}
\usepackage{amsthm}
\usepackage{mathtools}
\usepackage{bbm}

\newcommand{\defeq}{\coloneqq}

\DeclarePairedDelimiter{\prn}{\lparen}{\rparen}
\DeclarePairedDelimiter{\brk}{\lbrack}{\rbrack}
\DeclarePairedDelimiter{\brc}{\lbrace}{\rbrace}

\DeclareMathOperator{\Log}{Log}

\newcommand{\vect}[1]{\boldsymbol{#1}}

\DeclarePairedDelimiter{\norm}{\lVert}{\rVert}
\DeclarePairedDelimiterX{\inpr}[2]{\langle}{\rangle}{#1,#2}

\newcommand{\id}{\mathrm{id}}
\DeclareMathOperator{\diag}{diag}

\DeclareMathOperator{\tr}{Tr}
\DeclarePairedDelimiterX{\comm}[2]{\lbrack}{\rbrack}{#1,#2}